\documentstyle[preprint,eqsecnum,aps]{revtex}
\tightenlines                                                                                                                                                                

\newcommand{\be}{\begin{equation}}
\newcommand{\ee}{\end{equation}}
\newcommand{\bea}{\begin{eqnarray}}
\newcommand{\eea}{\end{eqnarray}}
\newcommand{\ba}{\begin{array}}
\newcommand{\ea}{\end{array}}

\newcommand{\ep}{\epsilon}
\newcommand{\Th}{\Theta}

\newcommand{\la}{\lambda}
\newcommand{\de}{\delta}
\newcommand{\dex}{\delta(x-y)}
\newcommand{\pa}{\partial}
\newcommand{\pax}{\partial_x}
\newcommand{\pari}{\partial^{-1}}
\newcommand{\no}{\nonumber}
\newcommand{\pade}{\pa_x\de (x-y)}
\newcommand{\epx}{\ep (x-y)}
\newcommand{\om}{\omega}
\newcommand{\tri}{\triangle}
\newcommand{\ub}{\underbrace}
\newcommand{\bo}{{\bf \mbox{\boldmath${\omega}$}\/}}
\begin{document}

\title{Nonlocal extended conformal algebras associated with multi-constraint 
KP hierarchy and their free-field realizations}

\author{Jiin-Chang Shaw$^1$, and Ming-Hsien Tu$^2$ }

\address{
$^1$ Department of Applied Mathematics, National Chiao Tung University, \\
Hsinchu, Taiwan, Republic of China,\\
and\\
$^2$ Department of Physics, National Tsing Hua University, \\
Hsinchu, Taiwan, Republic of China
}

\date{\today}

\maketitle

\begin{abstract}
We study the conformal properties of the multi-constraint
KP hierarchy and its nonstandard partner by covariantizing
their corresponding Lax operators. The associated 
second Hamiltonian structures
turn out to be nonlocal extension of $W_n$ algebra
by some integer or half-integer spin fields depending on
 the order of the Lax operators. In particular, we  show that the
complicated second Hamiltonian structure of the nonstandard
multi-constraint KP hierarchy can be simplified by
factorizing its Lax operator to multiplication form.
We then diagonalize this simplified Poisson matrix and obtain the
free-field realizations of its associated nonlocal algebras.
\end{abstract}
\newpage
%%%%%%%%%%%%%%%%%%%%%%%%%%%%%%%
\section{Introduction}
The relationship between conformal field theory and 
integrable systems has been paid much attention in the past few years. 
Especially many works are to 
explore the role played by classical extended conformal algebras in integrable systems 
\cite{DS,D,BS}. 
A typical example is the $n$-th Korteweg-de Vries (KdV) 
hierarchy \cite{D} which has Lax operator of the form
\be
l_n=\pa^n+u_2\pa^{n-2}+\cdots+u_n\qquad (n\geq 2)
\label{laxkdv}
\ee
and satisfies the hierarchy equations
\be
\pa_kl_n=[(l_n^{k/n})_+,l_n].
\label{eqkdv}
\ee
The second Hamiltonian structure of (\ref{laxkdv}) is just the second
Gelfand-Dickey (GD) bracket \cite{D} with Dirac constraint imposed by $u_1=0$ which, 
in operator form, reads
\be
\Th_2(\frac{\de H}{\de l_n})
=(l_n\frac{\de H}{\de l_n})_+l_n-
l_n(\frac{\de H}{\de l_n}l_n)_++\frac{1}{n}[l_n,\int^x res[l_n,\frac{\de H}{\de l_n}]]
\label{hskdv}
\ee
with
\be
\frac{\de H}{\de l_n}=\pa^{-n+1}\frac{\de H}{\de u_2}+
\cdots+\pa^{-1}\frac{\de H}{\de u_n}.
\label{varkdv}
\ee
where $(A)_{\pm}$ denote the differential part and the 
integral part of the pseudo-differential operator 
$A$ respectively.
From (\ref{hskdv}),  the classical Virasoro algebra can be constructed as the
Pisson bracket for $u_2\equiv t_2$ \cite{GN,K}, i.e.,
\be
\{t_2(x),t_2(y)\}=[\frac{n^3-n}{12}\pa_x^3+2t_2(x)\pa_x+t'_2(x)]\dex.
\label{viralg}
\ee
Therefore we can identify the coefficient function $u_2$ in $l_n$ as
the Virasoro generator $t_2$. Futhermore, the $W_n$ algebra \CITE{Z}, which is the
nonlinear extension of the Virasoro algebra, was also constructed as the
second GD bracket algebra for  higher spin fields $w_k (k\geq 3)$ 
which can be expressed in terms of the differential polynomial of the coefficient 
functions $\{u_2,u_3,\cdots,u_n\}$ in $l_n$ \cite{M,B,WPSK}. 
Recently, a systematic procedure 
for constructing these higher spin fields $w_k$ from the Lax operator $l_n$
has been developed by Di Francesco, Itzykson and Zuber (DIZ)\cite{DIZ}.
It not only  gives an efficient way to covariantize a (pseudo-)differential operator but also
provides a clear connections between classical extended conformal
algebras and the Hamiltonian structures of certain integrable systems.
In the following, let us briefly recall the DIZ's procedure for covariantizing
a differential operator.

A function $f(x)$ is called a spin-$k$ field if it transforms under 
coordinate change $x\rightarrow t(x)$ as $f(t)=(dx/dt)^kf(x)$. 
The space of all spin-$k$ fields is denoted by $F_k$. An operator
$\triangle$ is called a covariant operator if it maps from $F_h$
to $F_l$ for some $h$ and $l$. We denote this by 
$\triangle: F_h\rightarrow F_l$, i.e., under the change
$x\rightarrow t(x)$
\be
\tri(t)=(\frac{dx}{dt})^l\tri(x)(\frac{dx}{dt})^{-h}.
\label{covtran}
\ee
The infinitesimal change of (\ref{covtran})  under 
$t(x)=x-\ep(x)$ can be easily written down as
\be
\de_{\ep}\tri(x)=[\ep(x)\pa_x+l\ep'(x)]\tri(x)-\tri(x)[\ep(x)\pa_x+h\ep'(x)]
\label{infini}
\ee
which will be useful latter on. 

The essential concept for covariantizing the differential operator $l_n$ is that 
the Virasoro flow generated by $\int dx t_2(x)\ep(x)$ is just
the infinitesimal form (\ref{infini}). Thus, for a suitable choice of $h$ and $l$,
one can covariantize the differential operator $l_n$. It is not hard to show \cite{DIZ}
that
\be
l_n: F_{-\frac{n-1}{2}}\rightarrow F_{\frac{n+1}{2}}.
\label{mapkdv}
\ee
The remaining task is to construct a family of covariant operators such that
each of them maps from $F_{-\frac{n-1}{2}}$ to $F_{\frac{n+1}{2}}$ and 
depends on a spin field and the Virasoro generator $t_2(x)$. 
To this end, one can parametrize the Virasoro generator $t_2$ by
\be
t_2(x)=\frac{n^3-n}{12}(b'(x)-\frac{1}{2}b^2(x))
\label{paravir}
\ee 
where $b(x)$ is an anomalous spin-1 field which satisfies the transformation law:
$b(t)=(dx/dt)b(x)+(d^2x/dt^2)(dx/dt)^{-1}$. The usage of $b$ is to define a sequence
of covariant operators:
\be
D_k^l\equiv [\pa-(k+l-1)b][\pa-(k+l-2)b]\cdots[\pa-kb]\qquad (l\geq 0)
\ee
which maps from $F_k$ to $F_{k+l}$. Thus given a spin-$k(\geq 1)$ field $w_k$,
one can construct the following covariant operator
\be
\tri_k(w_k,t_2)\equiv \sum_{l=0}^{n-k}\alpha_{k,l}^{(n)}[D_k^lw_k]
D_{-\frac{n-1}{2}}^{n-k-l}
\label{covop}
\ee
such that $\tri_k(w_k,t_2): F_{-(n-1)/2}\rightarrow F_{(n+1)/2}$. Here the
coefficients 
\be
\alpha_{k,l}^{(n)}=
\frac
{\left(
\ba{c}
k+l-1\\ l
\ea
\right)
\left(
\ba{c}
n-k\\ l
\ea
\right)}
{\left(
\ba{c}
2k+l-1\\ l
\ea
\right)}
\ee
are determined by requiring the right-hand side of (\ref{covop}) to depend on $b$
only through $t_2$ defined by (\ref{paravir}). With these definitions in hand, one can
write down the covariant form of $l_n$ as \cite{DIZ}
\be
l_n=D^{n}_{-\frac{n-1}{2}}+\sum_{k=3}^n\triangle_k(w_k,t_2).
\label{covform}
\ee
Eq. (\ref{covform}) decomposes the coefficient functions $u_i$'s into the spin fields and the
Virasoro generator. Inverting these relations then gives the expressions for spin fields
as differential polynomials of $u_i$'s.  Thus, in terms of $t_2$ and the 
spin fields $w_k$, the second GD bracket defines the $W_n$-algebra.

Another remarkable property associated with the $n$-th KdV hierarchy is that
by factorizing the Lax operator $l_n$, the second Hamiltonian structure can be
transformed to a much simpler one in an appropriate space of the modified variables.
Such factorization not only provides a Miura transformation which maps
the $n$-th KdV hierarchy to the corresponding modified hierarchy, but also
gives a free-field realization of the $W_n$ algebra. This is what we called
the Kupershmidt-Wilson (KW) theorem \cite{KW,D2}.

The above scheme can be generalized to Kadomtsev-Petviashvili (KP) hierarchy 
and its reductions associated with appropriate pseudo-differential
operators \cite{FMR,DHP,R,DH,H,HSY}. Especially, the conformal properties of 
the constrained KP hierarchy \cite{KS,KSS,C1}  
and its nonstandard partner (also called constrained modified KP hierarchy \cite{OS})
have been considered in \cite{HSY}. The KW theorem for these hierarchies
are also investigated in \cite{C2,L2,ST}.
The purpose of this paper is to generalize the previous
work for the one-constraint KP hierarchy \cite{HSY}
to study the conformal properties of the multi-constraint KP hierarchy and
its nonstandard partner [for the definitions of these hierarchies, see Sec. II and III]. 
We will follow the approach in \cite{HSY} to covariantize their corresponding
Lax operators and obtain the associated algebras with respect to their
second Hamiltonian structures. In particular, we will study the 
KW theorem for the nonstandard multi-constraint KP hierarchy.

Our paper is organized as follow:
In Sec. II, we covariantize the Lax operator of the multi-constrant KP hierarchy.
We show that the associated second Hamiltonian structure defines a nonlocal
extension of  $W_n$ algebra by $2m$ spin-$\frac{n+1}{2}$ fields. We give an example
($n=3$) to illustrate such nonlocal algebra in detail. In Sec. III, 
we consider the nonstandard multi-constraint KP hierarchy and its associated algebras.
We investigate the conformal property of the nonstandard Lax operator
$K_{(n,m)}$ through the new operator
$\hat{L}_{(n+1,m-1)}\equiv\pa K_{(n,m)}$. We show that the second Hamiltonian structure 
defined by $K_{(n,m)}$ is mapped to the sum of the second and the third
GD brackets defined by $\hat{L}_{(n+1,m-1)}$. This helps us to 
covariantize the operator $\hat{L}_{(n+1,m-1)}$ and obtain the associated algebras.
In Sec. IV, We discuss the KW theorem (or Miura transformation) for the 
nonstandard multi-constraint KP hierarchy. Conclusion is presented in Sec. V.

%%%%%%%%%%%%%%%%%%%%%%%%%%%%%%%%
\section{The multi-constraint KP hierarchy}
The multi-constraint KP hierarchy \cite{OS} is the ordinary KP hierarchy
restricted to pseudo-differential operator of the form
\be
L_{(n,m)}=\pa ^n+u_2\pa^{n-2}+\cdots+u_n+
\sum_{i=1}^m\phi_i\pari\psi_i
\label{laxckp}
\ee
which satisfies the hierarchy equations
\bea
\pa_k L_{(n,m)}&=& [(L_{(n,m)}^{k/n})_+,L_{(n,m)}],
\label{eqckp}\\
\pa_k\phi_i &=&((L_{(n,m)}^{k/n})_+\phi_i)_0,
\qquad \pa_k\psi_i=-((L_{(n,m)}^{k/n})^*_+\psi_i)_0
\label{eqwf}
\eea
where $\phi_i$ and $\psi_i$ are eigenfunctions and adjoint eigenfunctions, respectively.
It is easy to show that (\ref{eqwf}) is consistent with (\ref{eqckp}).
(Notations:$(A)_0$ denotes the zeroth order term, and * stands for the conjugate
operation: $(AB)^*=B^*A^*,\; \partial^*=-\partial,\; f(x)^*=f(x)$). 
The second Hamiltonian structure associated with the Lax operator
(\ref{laxckp}) is given by \cite{OS}
\bea
\Th_2(\frac{\de H}{\de L_{(n,m)}})
&=&(L_{(n,m)}\frac{\de H}{\de L_{(n,m)}})_+L_{(n,m)}-
L_{(n,m)}(\frac{\de H}{\de L_{(n,m)}}L_{(n,m)})_+\no\\
& &+\frac{1}{n}[L_{(n,m)},\int^x res[L_{(n,m)},\frac{\de H}{\de L_{(n,m)}}]].
\label{hsckp}
\eea
where one can parametrize the differential of a Hamiltonian
$H$ by
\be
\frac{\de H}{\de L_{(n,m)}}
=\pa^{-n+1}\frac{\de H}{\de u_2}+\cdots+\pa^{-1}\frac{\de H}{\de u_n}+B
\ee
where $B=(B)_{\geq 0}$ is any differential operator satisfying
\be
(B\phi_i)_0=\frac{\de H}{\de \psi_i},\qquad (B^*\psi_i)_0=\frac{\de H}{\de \phi_i}
\ee
which can be easily obtained from the following relation
\be
\de H=\int res(\frac{\de H}{\de L_{(n,m)}}\de L_{(n,m)})=
\int [\sum_{i=2}^n\frac{\de H}{\de u_i}\de u_i+\sum_{i=1}^m
(\frac{\de H}{\de \phi_i}\de \phi_i+\frac{\de H}{\de \psi_i}\de \psi_i)].
\label{varh}
\ee

Eq. (\ref{hsckp}) contains nonlocal brackets, for example
\bea
\{\phi_i(x),\phi_j(y)\}&=&
-\frac{1}{n}\phi_i(x)\epx\phi_j(y)-\phi_j(x)\epx\phi_i(y),\no\\
\{\psi_i(x),\psi_j(y)\}&=&
-\frac{1}{n}\psi_i(x)\epx\psi_j(y)-\psi_j(x)\epx\psi_i(y),\\
\{\phi_i(x),\psi_j(y)\}&=&
\de_{ij}L_{(n,m)}\dex+\frac{1}{n}\phi_i(x)\epx\psi_j(y),\no
\eea
where $\epx\equiv \pax^{-1}\dex$ is the antisymmetric step function.
Based on the observation \cite{HSY} that 
$(L_{(n,m)})_+$ and $(L_{(n,m)})_-$ transform
independently under the change of coordinate,  the positive part
$(L_{(n,m)})_+$ has the same conformal property as the Lax operator $l_n$
of the $n$-th KdV hierarchy, and the negative part $(L_{(n,m)})_-$ then transforms
according to
\be
\phi_i(t)\pa_t^{-1}\psi_i(t)=(\frac{dx}{dt})^{\frac{n+1}{2}}
\phi_i(x)\pa_x^{-1}\psi_i(x)(\frac{dx}{dt})^{\frac{n-1}{2}}
\label{lax-}
\ee
such that the Lax operator $L_{(n,m)}$ transforms covariantly from
$F_{-\frac{n-1}{2}}$ to $F_{\frac{n+1}{2}}$. 
Eq. (\ref{lax-}) implies that
\bea
\phi_i(t)&=&(\frac{dx}{dt})^{\frac{n+1}{2}}\phi_i(x),
\label{phitf}\\
\psi_i(t)&=&(\frac{dx}{dt})^{\frac{n+1}{2}}\psi_i(x).
\label{psitf}
\eea
In other words, $\phi_i$ and $\psi_i$ are both spin-$\frac{n+1}{2}$ fields
and we denote them by 
$w_{\frac{n+1}{2}}^{(i)+}$ and $w_{\frac{n+1}{2}}^{(i)-}$, 
respectively.
From the above discussions, the covariant form of the Lax operator $L_{(n,m)}$
can be easily written as
\be
L_{(n,m)}=D^{n}_{-\frac{n-1}{2}}+\sum_{k=3}^n\triangle_k(w_k,t_2)+
\sum_{i=1}^m w_{\frac{n+1}{2}}^{(i)+}\pari w_{\frac{n+1}{2}}^{(i)-}
\ee
and the second structure (\ref{hsckp}),
when expressed in terms of $t_2$ and the spin fields $w_3,\cdots,w_n,
w_{\frac{n+1}{2}}^{(i)\pm}$, defines a nonlocal extension of $W_n$ algebra
by $2m$ spin-$\frac{n+1}{2}$ fields.

Let us consider an example with $n=3$. For this case, the covariant form
of the Lax operator reads
\be
L_{(3,m)}=\pa^3+t_2\pa+w_3+\frac{1}{2}t_2'+\sum_{i=1}^m
w_2^{(i)+}\pari w_2^{(i)-}.
\ee
From (\ref{hsckp}) the Poisson brackets between the variables 
$\{t_2,w_3,w_2^{(i)\pm}\}$ can be written down as follows:
\bea
\{t_2(x),t_2(y)\}&=&[2\pa_x^3+2t_2(x)\pa_x+t_2'(x)]\dex \no\\
\{w_3(x),t_2(y)\}&=&[3w_3(x)\pa_x+w_3'(x)]\dex,\no\\
\{w_2^{(i)\pm}(x),t_2(y)\}&=&[2w_2^{(i)\pm}(x)\pa_x+w_2^{(i)\pm'}(x)]\dex,\no\\
\{w_3(x),w_3(y)\}&=&-[\frac{1}{6}\pa_x^5+\frac{5}{6}t_2(x)\pa_x^3+
\frac{5}{4}t_2'(x)\pa_x^2\no\\
& &+(\frac{3}{4}t_2''(x)+\frac{2}{3}t_2^2(x))\pa_x+
\frac{1}{6}t_2'''(x)+\frac{2}{3}t_2(x)t_2'(x)]\dex\no\\
& &+[4\sum_{i=1}^mw_2^{(i)+}(x)w_2^{(i)-}(x)\pa_x+
2\sum_{i=1}^m(w_2^{(i)+}(x)w_2^{(i)-}(x))']\dex ,
\label{w3}\\
\{w_2^{(i)\pm}(x),w_3(y)\}&=&\pm[\frac{5}{3}w_2^{(i)\pm}(x)\pa_x^2
+\frac{5}{2}w_2^{(i)\pm'}(x)\pa_x+w_2^{(i)\pm''}(x)+
\frac{2}{3}t_2(x)w_2^{(i)\pm}(x)]\dex,\no\\
\{w_2^{(i)\pm}(x),w_2^{(j)\pm}(y)\}&=&
-w_2^{(j)\pm}(x)\epx w_2^{(i)\pm}(y)-
\frac{1}{3}w_2^{(i)\pm}(x)\epx w_2^{(j)\pm}(y),\no\\
\{w_2^{(i)+}(x) ,w_2^{(j)-}(y) \}&=&\de_{ij}[\pa_x^3+t_2(x)\pa_x+w_3(x)+
\frac{1}{2}t_2'(x)]\dex+\de_{ij}
\sum_{k=1}^mw_2^{(k)+}(x)\epx w_2^{(k)-}(y)
\no\\
&&+\frac{1}{3}w_2^{(i)+}(x)\epx w_2^{(j)-}(y).\no
\eea
where $i,j=1,2,\cdots,m$.
This is a nonlocal extension of $W_3$ algebra by $2m$ spin-2 fields
$w_2^{(i)\pm}$. As $m=1$, (\ref{w3}) reduces to the previous result \cite{HSY}.

%%%%%%%%%%%%%%%%%%%%%%%%%%%%%%%%%%
\section{ The nonstandard multi-constraint KP hierarchy}
Now let us performing the following gauge transformation to obtain
the nonstandard Lax operator 
\be
K_{(n,m)}=\phi_1^{-1}L_{(n,m)}\phi_1\equiv
\pa^n+v_1\pa^{n-1}+\cdots+\pari v_{n+1}+\sum_{i=1}^{m-1}q_i\pari r_i
\label{gt}
\ee
where
\bea
v_1&=&n\frac{\phi_1'}{\phi_1},\no\\
v_2&=&u_2+\frac{n(n-1)}{2}\frac{\phi_1''}{\phi_1},\no\\
&\cdots &
\label{utov}\\
v_{n+1}&=&\phi_1\psi_1,\no\\
q_i&=&\phi_1^{-1}\phi_{i+1},\qquad r_i=\phi_1\psi_{i+1}.\no
\eea
In terms of $K_{(n,m)}$, (\ref{eqckp}) now satisfies the 
nonstandard multi-constraint KP hierarchy \cite{OS}
\bea
\pa_kK_{(n,m)} &=& [(K_{(n,m)}^{k/n})_{\ge 1},K_{(n,m)}],\no\\
\pa_kq_i &=&((K_{(n,m)}^{k/n})_{\ge 1}q_i)_0,
\label{nhe}\\
\pa_kv_{n+1} &=&-((K_{(n,m)}^{k/n})^*_{\ge 1}v_{n+1})_0,\qquad 
\pa_kr_i=-((K_{(n,m)}^{k/n})^*_{\ge 1}r_i)_0.\no
\eea
The second Hamiltonian structure associated with (\ref{gt}) is given by\cite{OS}
\bea
\Th^{NS}_2(\frac{\de H}{\de K_{(n,m)}})
&=&(K_{(n,m)}\frac{\de H}{\de K_{(n,m)}})_+K_{(n,m)}-
K_{(n,m)}(\frac{\de H}{\de K_{(n,m)}}K_{(n,m)})_+
+[K_{(n,m)},(K_{(n,m)}\frac{\de H}{\de K_{(n,m)}})_0]\no\\
& &+\pari (res[K_{(n,m)},\frac{\de H}{\de K_{(n,m)}}])K_{(n,m)}
+[K_{(n,m)},\int^x res[K_{(n,m)},\frac{\de H}{\de K_{(n,m)}}]],
\label{hsnckp}
\eea
where
\be
\frac{\de H}{\de K_{(n,m)}}=\pa^{-n} \frac{\de H}{\de v_1}+\cdots+
\frac{\de H}{\de v_{n+1}}+A
\ee
with $A$ satisfying
\bea
(A)_0 &=&0,\no\\
(Aq_i)_0 &=&\frac{\de H}{\de r_i}-q_i\frac{\de H}{\de v_{n+1}},
\label{ida}\\
(A^*r_i)_0 &=&\frac{\de H}{\de q_i}-r_i\frac{\de H}{\de v_{n+1}}.\no
\eea
Now it is quite natural to ask: what about the conformal property
of $K_{(n,m)}$ and what kind of the algebra associated with
the Poisson bracket (\ref{hsnckp}). At first sight to the gauge transformation
(\ref{gt}) and to (\ref{mapkdv}) and (\ref{phitf}), one would expect that
\be
K_{(n,m)}: F_{-n}\rightarrow F_0
\label{mapnckp}
\ee
and the Virasoro generator is just the original one, i.e.,
\be
t_2\equiv u_2=v_2-\frac{n-1}{2}v_1'-\frac{n-1}{2n}v_1^2.
\label{oldvir}
\ee
However, it was pointed out \cite{HSY} that such identification is not correct.
This choice of the Virasoro generator will let $v_1$ not be a spin-1 field. 
To conquer this problem, we have to abandon (\ref{mapnckp})
and find a new Virasoro generator
such that (\ref{hsnckp}) gives an extended conformal algebra. 
Let us go back to the example for $n=3$. 
The nonstandard Lax operator $K_{(3,m)}$  then reads
\be
K_{(3,m)}=\pa^3+v_1\pa^2+v_2\pa+v_3+\pari v_4+\sum_{i=1}^{m-1}
q_i\pari r_i
\ee
Using (\ref{w3}) , (\ref{utov}) and (\ref{oldvir}), we have
\bea
\{v_1(x),v_1(y)\}&=&12\pade,\\
\{v_1(x),t_2(y)\}&=&[6\pa_x^2+v_1(x)\pa_x+v_1'(x)]\dex
\eea
It is obvious that $v_1$ is not a spin-1 field due to the anomalous
term  ``$6\pa_x^2$". However, if we define 
$t_2\equiv v_2-\frac{1}{2}v_1'-\frac{1}{3}v_1^2$ then
\bea
\{v_1(x),t_2(y)\}&=&[v_1(x)\pa_x+v_1'(x)]\dex,\\
\{t_2(x),t_2(y)\}&=&[5\pa_x^3+2t_2(x)\pa_x+t_2'(x)]\dex.
\eea
Therefore, with such new choice of the Virasoro generator, $v_1$ becomes 
a spin-1 field. Note that the coefficient of the anomalous term ``$\pa_x^3\dex$" 
in $\{t_2(x),t_2(y)\}$ is just given by $\frac{n(n^2-1)}{12}$ with $n=4$. On the
other hand, if we calculate the Poisson brackets for 
$\{q_i(x),t_2(y)\}$ and $\{r_i(x),t_2(y)\}$, we find
\bea
\{q_i(x),t_2(y)\}&=&\frac{5}{2}q_i'(x)\dex-\frac{3}{2}\epx q_i''(y),\\
\{r_i(x),t_2(y)\}&=&[\frac{5}{2}r_i\pa_x+r_i'(x)]\dex.
\eea
Here  we see that $r_i$ are spin-$\frac{5}{2}$ fields but $q_i$ are not spin fields.
However if we take the derivative of $q_i$ then $q_i'(x)$ become  spin-$\frac{5}{2}$
fields, i.e.,
\be
\{q'_i(x),t_2(y)\}=[\frac{5}{2}q'_i\pa_x+q''_i(x)]\dex.
\ee

Based on these observations, we are motivated to consider the operator
\be
\hat{L}_{(n+1,m-1)}=\pa K_{(n,m)}
\equiv \pa^{n+1}+\hat{u}_1\pa^n+\cdots+\hat{u}_{n+1}+
\sum_{i=1}^{m-1}\hat{\phi}_i\pari \hat{\psi}_i
\label{ktol}
\ee
where
\bea
\hat{u}_1&=&v_1,\no\\
\hat{u}_2&=&v_2+v_1',\no\\
& \cdots&\no\\
\hat{u}_n&=&v_n+v'_{n-1},
\label{vtou}\\
\hat{u}_{n+1}&=&v_{n+1}+v'_n+\sum_{i=1}^{m-1}q_ir_i,\no\\
\hat{\phi}_i&=&q'_i,\no\\
\hat{\psi}_i&=&r_i.\no
\eea
From (\ref{ida}) and (\ref{vtou}), we can show [see Appendix] that 
\be
\frac{\de H}{\de K_{(n,m)}}\pari=
\frac{\de H}{\de \hat{L}_{(n+1,m-1)}}+O(\pa^{-n-2}),
\label{vark}
\ee
which together with (\ref{ktol}) imply \cite{HSY} that 
the Hamiltonian structure associated with $\hat{L}_{(n+1,m-1)}$
is
\bea
\Omega(\frac{\de H}{\de \hat{L}_{(n+1,m-1)}})
&=&\pa\Th_2^{NS}(\frac{\de H}{\de K_{(n,m)}})\\
&=&(\hat{L}_{(n+1,m-1)}
\frac{\de H}{\de \hat{L}_{(n+1,m-1)}})_+\hat{L}_{(n+1,m-1)}-
\hat{L}_{(n+1,m-1)}(\frac{\de H}{\de \hat{L}_{(n+1,m-1)}}
\hat{L}_{(n+1,m-1)})_+\no\\
& &+[\hat{L}_{(n+1,m-1)},\int^x res[\hat{L}_{(n+1,m-1)},
\frac{\de H}{\de \hat{L}_{(n+1,m-1)}}]],
\label{hs23}
\eea
where the last term in (\ref{hs23}) is called the third GD structure which is 
compatible with the second GD stucture \cite{DIZ}. Therefore under the 
transformation (\ref{ktol}) the Hamiltonian structure (\ref{hsnckp}) is mapped to the sum of the 
second and the third GD brackets defined by the operator $\hat{L}_{(n+1,m-1)}$.
Now we can follow the DIZ's procedure to find a Virasoro generator such that
the Virasoro flow defined by (\ref{hs23}) gives
\be
\de_{\ep}\hat{L}_{(n+1,m-1)}=
[\ep(x)\pa_x+\frac{n+2}{2}\ep'(x)]\hat{L}_{(n+1,m-1)}-
\hat{L}_{(n+1,m-1)}[\ep(x)\pa_x-\frac{n}{2}\ep'(x)].
\ee
The Virasoro generator then is given by
\be
t_2=\hat{u}_2-\frac{n}{2}\hat{u}_1'-\frac{n-1}{2n}\hat{u}_1^2
\label{vir23}
\ee
which enables us to write down the covariant form of $\hat{L}_{(n+1,m-1)}$
as follows:
\be
\hat{L}_{(n+1,m-1)}=D^{n+1}_{-\frac{n}{2}}+\triangle_1(\hat{u}_1,t_2)+
\frac{n-1}{2n}\triangle_2(\hat{u}_1^2,t_2)+
\sum_{k=3}^{n+1}\triangle_k(\hat{w}_k,t_2)+
\sum_{i=1}^{m-1} \hat{w}_{\frac{n+2}{2}}^{(i)+}\pari \hat{w}_{\frac{n+2}{2}}^{(i)-}.
\label{cov23}
\ee
which maps from $F_{-\frac{n}{2}}$ to $F_{\frac{n+2}{2}}$.
Note that $\hat{\phi}_i\equiv\hat{w}_{\frac{n+2}{2}}^{(i)+}$ 
and $\hat{\psi}_i\equiv\hat{w}_{\frac{n+2}{2}}^{(i)-}$.
Since (\ref{cov23}) contains a spin-1 field $\hat{u}_1$ which satisfies
\bea
\{\hat{u}_1(x),\hat{u}_1(y)\}&=&n(n+1)\pade ,\\
\{\hat{u}_1(x),t_2(y)\}&=&[\hat{u}_1(x)\pa_x+\hat{u}_1'(x)]\dex,
\eea 
the associated algebra becomes the nonlocal extension of     
$U(1)$-$W_{n+1}$ algebra by $(2m-2)$
spin-$\frac{n+2}{2}$ fields.

%%%%%%%%%%%%%%%%%%%%%%%%%%%%%%%%%5
\section{The Miura transformation and free-field realization}
In this section, we will show that the Poisson structure defined by 
$\hat{L}_{(n+1,m-1)}$ can be simplified if we factorize $\hat{L}_{(n+1,m-1)}$
to multiplication form. In fact, $\hat{L}_{(n+1,m-1)}$ has 
the following representation \cite{ANPZ,A,Y}
\be
\hat{L}_{(n+1,m-1)}=(\pa-a_1)\cdots(\pa-a_p)(\pa-b_1)^{-1}\cdots(\pa-b_q)^{-1}
\label{facl}
\ee
where $p=n+m,q=m-1$. Comparing with (\ref{ktol}) and (\ref{facl}) 
we can obtain the relationship
between $\{\hat{u}_i,\hat{\phi}_i,\hat{\psi}_i\}$ and $\{a_i,b_i\}$ which is called 
the Miura transformation. Now let us first consider the second GD bracket 
(the first two terms of (\ref{hs23}))
under the factorization (\ref{facl}). 
Thanks to the generalized KW theorem \cite{ANPZ,A,Y,D3,MR}, the second 
GD bracket can be simplified to
\be
\{F,G\}_2^{GD}=-\sum_{i=1}^p\int (\frac{\de F}{\de a_i})(\frac{\de G}{\de a_i})'
+\sum_{i=1}^q\int (\frac{\de F}{\de b_i})(\frac{\de G}{\de b_i})'
\ee
which implies
\bea
\{a_i(x), a_j(y)\}_2^{GD} &=& -\de_{ij}\pade \no\\
\{b_i(x), b_j(y)\}_2^{GD} &=& +\de_{ij}\pade   
\label{pos2}\\
\{a_i(x), b_j(y)\}_2^{GD} &=&0 \no
\eea
On the other hand, it has been shown \cite{ST} that the third GD bracket
under the factorization (\ref{facl}) becomes
\be
\{F,G\}_3^{GD}=\int (\sum_{i=1}^p\frac{\de F}{\de a_i}+
\sum_{i=1}^q\frac{\de G}{\de b_i})(\sum_{j=1}^p\frac{\de F}{\de a_j}+
\sum_{j=1}^q\frac{\de G}{\de b_j})',
\ee
which leads to
\be
\{a_i(x),a_j(y)\}_3^{GD}=\{a_i(x),b_j(y)\}_3^{GD}=\{b_i(x),b_j(y)\}_3^{GD}=\pade.
\label{pos3}
\ee
Combining (\ref{pos2}) with (\ref{pos3}), we obtain
\bea
\{a_i(x), a_j(y)\} &=& (1-\de_{ij})\pade,\no\\
\{b_i(x), b_j(y)\} &=& (1+\de_{ij})\pade, 
\label{pos23}\\
\{a_i(x), b_j(y)\} &=&\pade. \no
\eea
Hence, for the Lax operator $K_{(n,m)}$, (\ref{pos23}) gives a simple and local
realization of its associated Poisson structure.
Especially, from (\ref{ktol}) and (\ref{facl}), the nonstandard Lax operator can be represented as
\be
K_{(n,m)}=\pari(\pa-a_1)\cdots(\pa-a_p)(\pa-b_1)^{-1}\cdots(\pa-b_q)^{-1}.
\label{simk}
\ee
In fact, we can go further to obtain the free-field realization
of the Poisson brackets (\ref{pos23}) and the nonstandard Lax operator (\ref{simk}). 
From (\ref{pos23}), the Poisson bracket matrix can be expressed as
\be
P_2={\bf h\/}{\bf h\/}^T-\hat{I}
\ee
where $T$ denotes the transpose operation and
\be
{\bf h\/}^T=(\ub{1,\cdots,1}_{p+q}), \qquad \hat{I}=
diag[\ub{1,\cdots,1}_p,\ub{-1,\cdots,-1}_q].
\ee
First, we can constructed $(p-1)$  orthonormal eigenvectors such that 
${\bf h\/}^T{\bf h\/}_i=0$ as follows:
\bea
{\bf h\/}_1^T&=&(1,-1,0,\cdots,0)/\sqrt{2}, \no\\
{\bf h\/}_2^T&=&(1,1,-2,0,\cdots,0)/\sqrt{6},\no\\
& &\cdots\no\\
{\bf h\/}_{p-1}^T&=&(1,1,\cdots,-p+1,0,\cdots,0)/\sqrt{p(p-1)}
\label{h+1}
\eea
It is easy to see that $P_2{\bf h\/}_i=-{\bf h\/}_i$. Hence these eigenvectors
 are degenerate with eigenvalue equals to $-1$. Similarly, we can also
construct another $(q-1)$  degenerate eigenvectors with eigenvalue equals to $+1$
as follows
\bea
{\bf h\/}_{p}^T&=&(0,\cdots,0,1,-1,0,\cdots,0)/\sqrt{2},\no\\
{\bf h\/}_{p+1}^T&=&(0,\cdots,0,1,1,-2,0,\cdots,0)/\sqrt{6},\no\\
&&\cdots \no\\
{\bf h\/}_{p+q-2}^T&=&(0,\cdots,0,1,1,\cdots,-q+1)/\sqrt{q(q-1)}.
\label{h-1}
\eea
It is obvious that these eigenvectors are orthogonal to each other.
Finally, from orthogonality, the remaining two orthonormal eigenvectors
have the following form
\be
{\bf h\/}_{\pm}^T=(1,1,\cdots,1,s_{\pm},\cdots,s_{\pm})/\sqrt{p+s_{\pm}^2q}.
\label{hpm}
\ee
Using $P_2{\bf h\/}_{\pm}=\la_{\pm} {\bf h\/}_{\pm}$, we have
\be
s_{\pm}=\frac{\la_{\pm}+1-p}{q},\qquad
\la_{\pm}=\frac{p+q\pm\sqrt{(p+q)^2-4(p-q-1)}}{2}.
\ee
From (\ref{h+1}), (\ref{h-1}) and (\ref{hpm}), 
the variable $\bo^T=(\om_1,\om_2,\cdots,\om_{p+q})
\equiv(a_1,\cdots,a_p,b_1,\cdots,b_q)$ can be expressed as
\be
\bo={\bf H\/}{\bf e\/}
\ee
where ${\bf H\/}$ is a $(p+q)\times(p+q)$ matrix defined by
\be
{\bf H\/}\equiv [{\bf h\/}_1,{\bf h\/}_2,\cdots,
{\bf h\/}_{p+q-2},{\bf h\/}_+,{\bf h\/}_-]
\ee
and ${\bf e\/}^T\equiv (e_1,e_2,\cdots,e_{p+q})$  satisfies
\be
\{e_i(x),e_j(y)\}=\la_i\de_{ij}\pade
\label{free}
\ee
with $\la_1=\cdots=\la_{p-1}=-1$, $\la_p=\cdots=\la_{p+q-2}=+1$,
$\la_{p+q-1}=\la_+$, and $\quad\la_{p+q}=\la_-$.
Therefore, when the spin fields $\{\hat{w}_k\}$ are expressed in $\{e_i\}$,
(\ref{free}) give a free-field realization of the associated nonlocal algebra
and the nonstandard Lax operator $K_{(n,m)}$ can be expressed
in terms of the free fields $\{e_i\}$ as 
\be
K_{(n,m)}=\pa^{-1}(\pa-({\bf H\/}{\bf e\/})_1)\cdots
(\pa-({\bf H\/}{\bf e\/})_p)(\pa-({\bf H\/}{\bf e\/})_{p+1})^{-1}
\cdots(\pa-({\bf H\/}{\bf e\/})_{p+q})^{-1}.
\ee
%%%%%%%%%%%%%%%%%%%
\section{Conclusions}
We have investigated the conformal property of the multi-constraint KP
hierarchy by covariantizing its Lax operator. We find that the result
is a straightforward generalization of the one-constraint ($m=1$) case \cite{HSY}.
The extended conformal algebras associated with (\ref{hsckp}) are nonlocal
extension of $W_n$ algebra by $2m$ spin-$\frac{n+1}{2}$ fields. 
We have presented the $n=3$ case to illustrate this nonlocal algebra explicitly.

We also study the nonstandard Lax operator which is obtained from the
Lax operator of the multi-constraint KP hierarchy by gauge transformation.
Because we can not covariantize the nonstandard Lax operator $K_{(n,m)}$
directly, we are motivated to covariantize a new operator 
$\hat{L}_{(n+1,m-1)}$ defined by (\ref{ktol}).
This new operator is still a pseudo-differential operator which is different
from the one-constraint case \cite{HSY} where the new operator is a differential one.
However, as we showed in Appendix that  Eq.(\ref{vark}), which was derived for the
one-constraint case, still holds for the present case.
Hence the second structure defined by the nonstandard Lax operator
can be mapped to the sum of the second and the third GD brackets defined by the
new operator. We have covariantized this new operator without difficulty and
obtained its associated extended conformal algebra. Furthermore, we find that
the sum of the second and the third GD  brackets can be simplified by factorizing
the new operator into multiplication form. We diagonalize this simplified Poisson
matrix to obtain the free-field realization of the associated nonlocal algebra and the
nonstandard Lax operator itself. 
Finally, we would like to remark that the matrix formulation of the
second GD bracket also gives rise to the nonlocal extension of the $W_n$-algebras
, called $V$-algebras \cite{Bil}.
It is interesting to explore the relationship between this matrix generalization
and the results presented here. Work in this direction is in progress.

{\bf Acknowledgments\/}
We would like to thank Professor W-J Huang 
for stimulating discussions. 
This work is supported by the National Science Council of the
Republic of China under grant No. NSC87-2811-M-007-0025.

%%%%%%%%%%%%%%%
\appendix
\section{}
From (\ref{vtou}), we have
\bea
\frac{\de H}{\de v_1}&=&\frac{\de H}{\de \hat{u}_1}-(\frac{\de H}{\de \hat{u}_2})'
\cdots-(-1)^n(\frac{\de H}{\de \hat{u}_n})^{(n)},\no\\
&\cdots&\no\\
\frac{\de H}{\de v_n}&=&\frac{\de H}{\de \hat{u}_n}-
(\frac{\de H}{\de \hat{u}_{n+1}})',\no\\
\frac{\de H}{\de v_{n+1}}&=&\frac{\de H}{\de \hat{u}_{n+1}},
\label{a1}\\
\frac{\de H}{\de q_i}&=&r_i\frac{\de H}{\de \hat{u}_{n+1}}-
(\frac{\de H}{\de \hat{\phi}_i})',\no\\
\frac{\de H}{\de r_i}&=&q_i\frac{\de H}{\de \hat{u}_{n+1}}+
(\frac{\de H}{\de \hat{\psi}_i}).\no
\eea
Then using
\be
f\pari=\pari f+\pa^{-2}f'+\pa^{-2}f''+\cdots
\ee
and (\ref{a1}) we have
\bea
\frac{\de H}{\de K_{(n,m)}}\pari
&=&[\pa^{-n}\frac{\de H}{\de v_1}+\cdots+
\frac{\de H}{\de v_{n+1}}]\pari,\no\\
&=&\pari\frac{\de H}{\de v_{n+1}}+\pa^{-2}
[\frac{\de H}{\de v_n}+(\frac{\de H}{\de v_{n+1}})']+\cdots\no\\
&&+\pa^{-n-1}[\frac{\de H}{\de v_1}+(\frac{\de H}{\de v_2})'
+\cdots+(\frac{\de H}{\de v_{n+1}})^{(n)}]+A\pari+O(\pa^{-n-2}),\no\\
&=&\pari\frac{\de H}{\de \hat{u}_{n+1}}+\pa^{-2}\frac{\de H}{\de \hat{u}_n}+
\cdots+\pa^{-n-1}\frac{\de H}{\de \hat{u}_1}+A\pari+O(\pa^{-n-2}),\no\\
&=&(\frac{\de H}{\de \hat{L}_{(n+1,m-1)}})_-+B+O(\pa^{-n-2}).
\eea
where $B\equiv A\pari$.
From (\ref{ida}) and (\ref{a1}) we have
\bea
(B\hat{\phi}_i)_0&=&-(A\pari\hat{\phi}_i)_0=(Aq_i)_0=\frac{\de H}{\de \hat{\psi}_i},\\
(B^*\hat{\psi}_i)_0&=&-(\pari A^*\hat{\psi})_0=
-(\pari A^*r_i)_0=\frac{\de H}{\de \hat{\phi}_i}
\eea
which imply that
\be
B=(\frac{\de H}{\de \hat{L}_{(n+1,m-1)}})_+
\ee
and hence
\be
\frac{\de H}{\de K_{(n,m)}}\pari=\frac{\de H}{\de \hat{L}_{(n+1,m-1)}}+O(\pa^{-n-2}).
\ee

\end{document}